\documentclass[11pt]{article}

\usepackage{a4}
\usepackage{color}
\usepackage{amssymb}
\usepackage[ruled,linesnumbered,algo2e,vlined]{algorithm2e}
\usepackage{amsmath}
\usepackage{amsmath,amssymb}
\usepackage{microtype}
\usepackage{hyperref}
\usepackage{wrapfig}
\usepackage{amsthm}
\usepackage{wrapfig}
\usepackage{lineno}
\usepackage{thm-restate}
\usepackage[T1]{fontenc}
\usepackage[utf8]{inputenc}
\usepackage[capitalise]{cleveref}
\usepackage{booktabs}
\usepackage{siunitx}

\usepackage[numbers,sort&compress]{natbib}
\usepackage[top=2cm, bottom=2cm, left=2cm, right=2cm, includefoot]{geometry}
\usepackage{authblk}
\usepackage{graphicx}
\usepackage{amsthm}
\usepackage{url}

\newtheorem{theorem}{Theorem}
\newtheorem{lemma}{Lemma}
\newtheorem{example}{Example}
\newtheorem{definition}{Definition}

\newtheorem{fact}[theorem]{Fact}

\SetCommentSty{mycommfont}

\SetKwProg{Fn}{Function}{}{end}

\newcommand{\mtt}[1]{\ensuremath{\mathtt{#1}}}
\newcommand{\GF}{\mathsf{GF}}
\newcommand{\Fact}{\mathsf{Fact}}
\newcommand{\GFact}{\mathsf{GaloisFactorization}}
\newcommand{\GRot}{\mathsf{GaloisRotation}}
\newcommand{\SPref}{\mathsf{SPref}}
\newcommand{\IsGal}{\mathsf{IsGalois}}
\newcommand{\OddPer}{\mathsf{Per}_o}
\newcommand{\EvenPer}{\mathsf{Per}_e}
\newcommand{\Per}{\mathsf{Per}}

\begin{document}

\title{Algorithms for Galois Words: Detection, Factorization, and Rotation}

\author[1]{Diptarama~Hendrian}
\author[2]{Dominik~K\"{o}ppl}
\author[3]{Ryo~Yoshinaka}
\author[3]{Ayumi~Shinohara}

\affil[1]{Tokyo Medical and Dental University, Japan}
\affil[2]{University of Yamanashi, Japan}
\affil[3]{Tohoku University, Japan}

\date{}

\maketitle            

\begin{abstract}
Lyndon words are extensively studied in combinatorics on words --- they play a crucial role on upper bounding the number of runs a word can have [Bannai+, SIAM J. Comput.'17].
We can determine Lyndon words, factorize a word into Lyndon words in lexicographically non-increasing order, and find the Lyndon rotation of a word, all in linear time within constant additional working space.
A recent research interest emerged from the question of what happens when we change the lexicographic order, which is at the heart of the definition of Lyndon words.
In particular, the alternating order, where the order of all odd positions becomes reversed, has been recently proposed.
While a Lyndon word is, among all its cyclic rotations, the smallest one with respect to the lexicographic order,
a Galois word exhibits the same property by exchanging the lexicographic order with the alternating order.
Unfortunately, this exchange has a large impact on the properties Galois words exhibit, which makes it a nontrivial task to translate results from Lyndon words to Galois words.
Up until now, it has only been conjectured that linear-time algorithms with constant additional working space in the spirit of Duval's algorithm are possible for computing the Galois factorization or the Galois rotation.

Here, we affirm this conjecture as follows.
Given a word $T$ of length $n$, we can determine whether $T$ is a Galois word, in $O(n)$ time with constant additional working space.
Within the same complexities, 
we can also determine the Galois rotation of $T$, 
and compute the Galois factorization of $T$ online.
The last result settles Open Problem~1 in [Dolce et al., TCS 2019] for Galois words.
\end{abstract}

\section{Introduction}
A \emph{Galois word} is a word that is strictly smaller than all its cyclic rotations with respect to the so-called alternating order,
where symbols at odd positions are compared in the usual lexicographic order, but symbols at the remaining positions in the opposite order.
While \mtt{aab} is clearly the smallest word among all its cyclic rotations \mtt{aba} and \mtt{baa} under the lexicographic order, 
\mtt{aab} is larger than \mtt{aba} under the alternating order because the \mtt{b} in the second position is smaller than \mtt{a}.
In fact, \mtt{aba} is a Galois word.
Readers familiar with Lyndon words may identify \mtt{aab} to be, nevertheless, a Lyndon word because it is strictly smaller than all its cyclic rotations with respect to the \emph{lexicographic} order.
While the definition of Lyndon and Galois words only differ by the used order, 
the combinatorial differences are astonishing.
For instance, on the one hand, Lyndon words cannot have proper borders, i.e., factors appearing both as a prefix \emph{and} as a suffix (but shorter than the word itself).
On the other hand, Galois words such as \mtt{aba} can have proper borders of odd lengths~\cite[Proposition~3.1]{reutenauer05mots}.

The name \emph{Galois word} has been coined by Reutenauer~\cite{reutenauer05mots}, who introduced these words and derived the naming by a bijection of Galois words and homographic classes of Galois numbers.
In the same paper~\cite{reutenauer05mots},  Reutenauer defined a unique factorization of a generalization of Lyndon words, a class of words covering Galois words.
Here, we call this factorization \emph{Galois factorization} since we only cover Galois words within the scope of this paper.
The Galois factorization is a factorization of a word into a sequence of non-increasing Galois words.
Later, Dolce et al.~\cite{dolce19generalized} could characterize the first factor of the Galois factorization~\cite[Theorem~33]{dolce19generalized},
and also provide a characterization of Galois words by their prefixes~\cite[Theorem~32]{dolce19generalized}.
However, Dolce et al. left it as an open problem (\cite[Open~Problem~1]{dolce19generalized}) to find a computation algorithm similar to Duval's algorithm~\cite{duval83lyndon} computing the Lyndon factorization.
In this paper, we solve this problem by introducing a factorization algorithm (\cref{alg:galoisfactorization} and \cref{thmGaloisFactorization}) in the spirit of Duval's algorithm, 
computing the Galois factorization in linear time with constant additional working space online.

Asides from the above results, we are only aware of the following two articles dealing with Galois words.
First, Dolce et al.~\cite{dolce19variations} studied generalized Lyndon-words, among them also Galois words, with respect to infinite orders.
Finally, Burcroff and Winsor~\cite{burcroff20generalized} gave a characterization of infinite generalized Lyndon words, and properties of how finite generalized Lyndon words can be infinitely extended.

\section{Related Work}
While we covered, to the best of our knowledge, all published results explicitly dealing with Galois words above,
Galois words have a strong relation with Lyndon words and the alternating order.

\subparagraph{Lyndon}
Regarding the former, an exhaustive list of results would go beyond the scope of this paper.
We nevertheless highlight that the Lyndon factorization (the aforementioned factorization when outputting factors that are Lyndon words) can be computed in linear time with constant additional space with Duval's algorithm~\cite{duval83lyndon}.
The same algorithm allows us to judge whether a word is Lyndon. 
Shiloach~\cite{shiloach81fast} gave a linear-time algorithm computing the Lyndon rotation of a primitive word $T$, i.e., its cyclic rotation that is Lyndon, in constant additional working space.

\subparagraph{Alternating Order}
Regarding the latter, much work focused on implementing a Burrows--Wheeler transform (BWT)~\cite{burrows94bwt} based on the alternating order.
While the classic BWT sorts all cyclic rotations of a given input word in lexicographic order, 
the alternating BWT~\cite{gessel12bijection} sorts the cyclic rotations with respect to the alternating order.
Gessel et al.~\cite{gessel12bijection} not only introduced this BWT variant, but also gave an inversion to restore the original word.
Subsequently, Giancarlo et al.~\cite{giancarlo20alternating} gave a linear-time algorithm for computing the alternating BWT\@.
To this end, they compute the Galois rotation of the input word $T$, i.e., the cyclic rotation of $T$ that is Galois.
However, their algorithm computing the Galois rotation needs an auxiliary integer array of length $n$.
Compared to space-efficient algorithms computing the classic BWT (e.g.~\cite{munro17cst} with linear time and space linear in the \emph{bit size} of the input text), 
this is a major bottleneck, but a necessary precomputation step of their algorithm constructing the alternating BWT\@.
Giancarlo et al.~\cite{giancarlo20alternating} also showed how to invert the alternating BWT in linear time.
In a follow-up~\cite{giancarlo23new}, Giancarlo et al. put their discovered properties of the alternating BWT into larger context by covering BWT variants based on a generalized ordering.
In that article, they also showed that the alternating BWT can be turned into a compressed self-index
	that supports pattern counting queries in time linear in the pattern length.
    The space of the index can be related with the number of symbol runs even when augmented for queries to locate all pattern occurrences in the text,
    by adapting r-indexing~\cite{gagie18bwt} techniques to the alternating BWT.

\subparagraph*{Our Contribution}
This paper makes three contributions to the research on Galois words.
First, we give an algorithm (\cref{thmCheckGalois} and \cref{alg:isgalois}) in \cref{secDeterminingGaloisWords} that 
checks, for a given word of length $n$, whether this word is Galois, in $O(n)$ time with constant additional working space.
Second, we give an algorithm (\cref{thmGaloisFactorization} and \cref{alg:galoisfactorization}) in \cref{secGaloisFactorization} that 
computes the Galois factorization in $O(n)$ time with constant additional working space online.
Finally, we show how to find the Galois rotation (\cref{thm:galoisRotation} and \cref{alg:galoisRotation}) in \cref{secGaloisRotation} that 
in $O(n)$ time with constant additional working space online, paving the way for constructing the alternating BWT in $o(n)$ working space.

We stress that, having an efficient Galois factorization algorithm allows us to merge indexing techniques for the alternate BWT with the bijective BWT~\cite{gil12bbwt,bannai19bbwt}
to give rise to a BWT-variant that indexes Galois words, whose indexing capabilities are left as future work.

\section{Preliminaries}

\newcommand*{\AltPrec}{\ensuremath{\prec_{\text{alt}}}}
\newcommand*{\AltPrecEq}{\ensuremath{\preceq_{\text{alt}}}}
\newcommand*{\AltSucc}{\ensuremath{\succ_{\text{alt}}}}
\newcommand*{\AltSuccEq}{\ensuremath{\succeq_{\text{alt}}}}
\newcommand*{\AltPrecM}{\ensuremath{\sqsubset_{\text{alt}}}}
\newcommand*{\LexPrec}{\ensuremath{\prec_{\text{lex}}}}
\newcommand*{\AltEq}{\ensuremath{=_{\text{alt}}}}

Let $\Sigma$ be a set of symbols called an \emph{alphabet}.
The set of words over $\Sigma$ is denoted by $\Sigma^*$.
The empty word is denoted by $\varepsilon$.
The \emph{length} of a word $W \in \Sigma^*$ is denoted by $|W|$.
The $i$-th symbol of a word $W$ is denoted by $W[i]$ for $1 \leq i \leq |W|$
and the factor of $W$ that begins at position $i$ and ends at position $j$ 
is $W[i..j]$ for $1 \leq i \leq j \leq |W|$.
We define $W[i..j] = \varepsilon$ if $i > j$.
A word $B$ is a \emph{border} of $W$ if $B$ is a prefix and a suffix of $W$.
We say a border $B$ of $W$ is \emph{proper} if $B \ne W$.
For a word $T$ we call an integer $p \in [1..|T|]$ a \emph{period} of $T$ if $T[i+p] = T[i]$ for all $i \in [1..|T|-p]$. 
In particular, $|T|$ is always a period of $T$.
Let $\OddPer(W)$ and $\EvenPer(W)$ be the shortest odd and even period of $W$ if any, respectively.
We set $\OddPer(W) = |W| + 1$ or $\EvenPer(W) = |W| + 1$ if $W$ does not have an odd or an even period, respectively.
Since the length of a word itself is a period,
a word of odd length always has an odd period and a word of even length always has an even period.
For a rational number $\alpha$, let $W^\alpha$ be the word obtained by concatenating $W$ $\alpha$ times.
Let $W^\omega$ be the infinite repetition of $W$.
We call a word $V \in \Sigma^*$ \emph{primitive} if the fact $V = U^k$ for some word $U \in \Sigma^*$ and an integer $k \ge 1$ implies $V = U$ and $k = 1$.
We say that two words $X$ and $Y$ have the same \emph{length-parity} if $|X| \mod 2 = |Y| \mod 2$, i.e., their lengths are both either odd or even.

We denote the standard lexicographic order over words with $\LexPrec$.
We define the \emph{alternating order} on words as follows:
Given two distinct words $S$ and $T$ such that $S^\omega \ne T^\omega$,
with the first mismatching symbol pair at a position $j$,
i.e., $S^\omega[1..j-1] = T^\omega[1..j-1]$ and $S^\omega[j] \neq T^\omega[j]$,
we write $S \AltPrec T$ if either (a) $j$ is odd and $S^\omega[j] < T^\omega[j]$, or (b) $j$ is even and  $S^\omega[j] > T^\omega[j]$.
In addition we denote by $S \AltEq T$ if $S^\omega = T^\omega$.
For instance, $\mtt{aba} \AltPrec \mtt{aab}$ but $\mtt{aab} \LexPrec \mtt{aba}$; 
$\mtt{b} \LexPrec \mtt{bba}$ but $\mtt{bba} \AltPrec \mtt{b}$; $\mtt{aba} \AltEq \mtt{abaaba}$.
We define $\varepsilon \AltSucc X$ for all $X \in \Sigma^+$.
We denote by $S \AltPrecEq T$ if either $S \AltPrec T$ or $S \AltEq T$.
We further write $S \AltPrecM T$ if $S \AltPrec T$ but neither $S$ is a prefix of $T$ nor vice versa.

We introduce $S \AltPrecM T$ for the following reason: 
For two words $S$ and $T$ with $S \AltPrec T$, it is generally not true that $S U \AltPrec T U$ for all words $U$ (e.g., $\mtt{ab} \AltPrec \mtt{aba}$ but $\mtt{abac} \AltPrec \mtt{abc}$).
However, for $\AltPrecM$ we have:
\begin{fact}\label{factAltPrecMExpansion}
For two words $S$ and $T$ with $S \AltPrecM T$, it holds that $S U \AltPrecM T U$ for all words $U$.
\end{fact}

We also make use of the following additional facts:
\begin{fact}\label{factSameLengthOmegaOrder}
    For two words $S$ and $T$ with $S^\omega = T^\omega$, there exists a primitive word $U$ integers $a$ and $b$ such that $S = U^a$ and $T = U^b$.
\end{fact}

\begin{fact}\label{factPeriodPrimitive}
    $T$ is non-primitive if and only if $|T|/p$ is an integer of at least two for $p$ being $T$'s smallest period.
    If $\Per_o(T) = |T|$, then $T$ is primitive.
\end{fact}

\begin{example}
We cannot switch $\Per_o$ with $\Per_e$ in Fact~\ref{factPeriodPrimitive}. A counter-example is 
the non-primitive word $T_1 = \mtt{abaaba}$, for which we have $\Per_o(T_1) = 3$ but $\Per_e(T_1) = 6$.
Also, for $T_2 = \mtt{aa}$ we have $\Per_o(T_2) = 1$ but $\Per_e(T_2) = 2$.
\end{example}

The following property holds for any two periods of a word.
\begin{lemma}[\cite{fine65uniqueness}]\label{lem:periodicityLemma}
    Let $p$ and $q$ be periods of a word $T$.
    If $p + q - r \le |T|$, then $r$ is also a period of $T$, where $r$ is the greatest common divisor of $p$ and $q$.
\end{lemma}

A word is called \emph{Galois} if it is, among all its cyclic rotations, the smallest with respect to $\AltPrec$.
By definition, a Galois word has to be primitive (otherwise, it has an identical cyclic rotation that is not strictly larger).
The following properties hold of Galois words.
\begin{lemma}[{\cite[Theorem 14]{dolce19generalized}}]\label{lem:GMRRS_7.4}
A primitive word $T$ is Galois if and only if $T$ is smaller than all its suffixes, with respect to $\AltPrec$.
\end{lemma}
\begin{lemma}[{\cite[Theorem 32]{dolce19generalized}}]\label{lem:dolceGalois}
    A word $T$ is Galois if and only if for any factorization $T = UV$ with $U,V \in\Sigma^+$, one of the following condition holds: 
    (1) $U \AltPrec T$ if $|U|$ is even; (2) $U \AltSucc T$ if $|U|$ is odd.
\end{lemma}
\begin{lemma}[{\cite[Lemma 35]{dolce19generalized}, \cite[Proposition~3.1]{reutenauer05mots}}]\label{lem:OddBorder}
If a Galois word $T$ has a proper border $B$, then the length of $B$ is odd.
\end{lemma}

For example \mtt{aba} and \mtt{abba} are Galois words with a proper border.
Unlike for Lyndon words (cf.~\cite[Proposition~1.3]{duval83lyndon}), 
it does not hold that, if $U$ and $V$ are Galois words then $UV$ is Galois if $U \AltPrec V$.
For instance, $\mtt{aba} \AltPrec \mtt{c}$ but $\mtt{abac}$ is not Galois because $\mtt{ac} \AltPrec \mtt{abac}$.

\section{Characteristics of Pre-Galois Words}
In this section, we define pre-Galois words and study their properties. 
The observations we make here will lead us to helpful tools that we will leverage for proposing the three algorithms in the subsequent sections, namely 1. determining Galois words, 2. computing the Galois factorization, and 3. computing the Galois rotation of a word.

\begin{definition}[Pre-Galois word]
    A word $T$ is a \emph{pre-Galois} word if every proper suffix $S$ of $T$ satisfies one or both of the following conditions: (1) $S$ is a prefix of $T$; (2) $S \AltSucc T$.
\end{definition}

In particular, a Galois word is pre-Galois.
However, the converse is in general not true;
for example $T = \mathtt{abaab}$ is pre-Galois but not Galois because $\mathtt{ab} \AltPrec \mathtt{abaab}$.
In what follows, we introduce a basic property of pre-Galois words.
\begin{lemma}
\label{lem:prepref}
    Let $U$ be a word that is not pre-Galois.
    Then, for any word $V$, $UV$ is not pre-Galois.
    The contraposition is that any prefix of a pre-Galois word is pre-Galois.
\end{lemma}
\begin{proof}
    By definition there exists a proper suffix $S$ of $U$ such that $S \AltPrecM U$.
    By \cref{factAltPrecMExpansion}, $S\cdot V \AltPrecM U\cdot V$ holds.
\end{proof}

\begin{figure}[t]
    \begin{minipage}{0.48\linewidth}
        \includegraphics{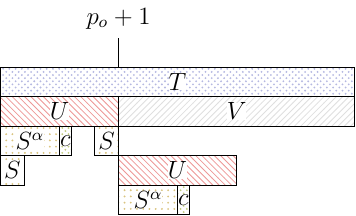}
    \end{minipage}
    \hfill
    \begin{minipage}{0.51\linewidth}
    \includegraphics{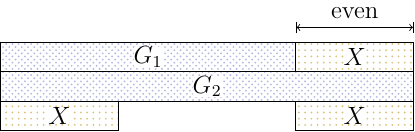}
    \end{minipage}
    \caption{Left: Sketch of the proofs of \cref{lem:PreGaloisPerOdd} and \cref{lem:PreGaloisPerEven}.
        Right: Sketch of the proof of \cref{lemGaloisRoot}. As both Galois roots are prefixes of $T$, we obtain a border $X$ of $G_2$ with even length which contradicts \Cref{lem:OddBorder}.
    }
    \label{fig:PreGaloisPerOdd}
\end{figure}

Next, we study properties of periods of pre-Galois words.
\begin{lemma}\label{lem:PreGaloisPerOdd}
Let $T$ be a pre-Galois word that has an odd period.
Let $p_o = \OddPer(T)$ be the shortest odd period of $T$.
Then $T[1..p_o]$ is Galois.
\end{lemma}
\begin{proof}
    Let $U = T[1..p_o]$ and $T = UV$.
    By \Cref{lem:prepref}, $U$ is pre-Galois.
    Assume $U$ is not Galois.
    Then there exists a proper suffix $S$ of $U$ such that $S$ is a prefix of $U$ and $S \AltPrecEq U$.
    Since $p_o$ is odd and the shortest odd period of $T$,
    $U$ is primitive according to \cref{factPeriodPrimitive}, and we obtain two observations: 
    First, by \cref{factSameLengthOmegaOrder}, if $S^\omega = U^\omega$ then $S = U$, a contradiction to $S$ being proper.
    Hence, $S \AltPrec U$ must hold.
    
    Second, there exists a rational number $\alpha \ge 1$ and a symbol $c \in \Sigma$
    such that $S^\alpha c$ is a prefix of $U$, $S^\omega[1..|S^\alpha c|] \AltPrec S^\alpha c$, and $|S^\alpha c| < |U|$.
    See the left of \cref{fig:PreGaloisPerOdd} for a sketch.
    By definition, we have $S^{\alpha-1} = U[1..|S^{\alpha-1}|]$.
    If $|S|$ is odd, we have $T[|S|+1..|T|] \AltPrecM T$, which implies that $T$ is not pre-Galois.
    Otherwise, if $|S|$ is even, $|U| + |S^{\alpha-1}c| \le |T|$ since $p_o$ is the shortest odd period of $T$ and therefore $|T| \ge 2|U|$ with $|U| \ge |S^{\alpha-1}c|$.
    Here, we have $T[|U|-|S|+1..|U|+|S^{\alpha-1}c|] = S^\omega[1..|S^{\alpha}c|]$.
    Therefore, $T[|U|-|S|+1..|U|+|S^{\alpha-1}c|] \AltPrecM T[1..|S^{\alpha}c|]$, which implies $T$ is not pre-Galois.
\end{proof}
A similar property also holds for even periods.
\begin{lemma}
\label{lem:PreGaloisPerEven}
Let $T$ be a pre-Galois word that has an even period.
Let $p_e = \EvenPer(T)$ be the shortest even period of $T$.
Then $T[1..p_e]$ is Galois if primitive.
\end{lemma}
\begin{proof}
    We follow the proof of \cref{lem:PreGaloisPerOdd} by replacing $U$ there with $T[1..p_e]$.
    We also give there $p_e$ the role of $p_o$.
    We can do that because we assume that $U$ is primitive, so we obtain a proper border $S$ of $U$ like in the previous proof.
\end{proof}

We are in particular interested in prefixes of pre-Galois words that are Galois.
To formalize this idea, we define \emph{Galois roots} of a pre-Galois word as follows.
\begin{definition}[Galois root]
    Let $P$ be a prefix of a pre-Galois word $T$.
    We call $P$ a \emph{Galois root} of $T$ if $|P|$ is a period of $T$ and $P$ is Galois.
\end{definition}
In addition to our aforementioned example $T = \mathtt{abaab}$, \mtt{aba} is a Galois root of $T$.
Also, the words $T[1..p_o]$ in \Cref{lem:PreGaloisPerOdd} and $T[1..p_e]$ \Cref{lem:PreGaloisPerEven}, are Galois roots of $T$ if they are, respectively, primitive.
Note that a pre-Galois word $T$ has at least one Galois root, namely $T$'s prefix of length equal to $T$'s shortest period.
While a pre-Lyndon word has exactly one Lyndon root, a pre-Galois word can have two different Galois roots:

\begin{lemma}
\label{lemGaloisRoot}
A pre-Galois word $T$ can have at most two Galois roots, and their lengths have different parities.
\end{lemma}
\begin{proof}
	Assume that there are two Galois roots~$G_1$ and $G_2$ with the same length-parity.
	Then the length difference of $G_1$ and $G_2$ is even.
	Without loss of generality, suppose that $|G_1| < |G_2|$. Then the suffix $X = G_2[|G_1|+1..]$ of $G_2$ is also a prefix of $G_2$ since $T = G_1^\alpha = G_2^\beta$ for rational numbers $\alpha$ and $\beta$.
	Hence, $X$ is a border of $G_2$ with even length, which is impossible due to \cref{lem:OddBorder}.
	See the right of \cref{fig:PreGaloisPerOdd} for a sketch.
\end{proof}

\newcommand*{\Godd}{\ensuremath{G_{\text{o}}}}
\newcommand*{\Geven}{\ensuremath{G_{\text{e}}}}
In what follows, we name the odd-length and the even-length Galois root, if they exist, by $\Godd$ and $\Geven$, respectively.
By \cref{lemGaloisRoot}, they are well-defined.
For example, consider $T=\mtt{aba}$.
The two prefixes $\mtt{ab}$ and $\mtt{aba}$ are both Galois, for which $T = (\mtt{ab})^{3/2} = (\mtt{aba})^1$.

From \Cref{lem:PreGaloisPerEven}, $T[1..p_e]$ is Galois only when it is primitive.
Next, we consider the case where $T[1..p_e]$ is not primitive.

\begin{lemma}
\label{lemGoddPrimeFull}
    Let $T$ be an even-length pre-Galois word with no even-length Galois root, i.e., $T[1..p_e]$ is not primitive, where $p_e = \EvenPer(T)$.
	Then there exists $\Godd = T[1..p_o]$ such that $T = \Godd^{k}\Godd'$ with $k \ge 2$,
    where $p_o = \OddPer(T)$ and $\Godd'$ is a prefix of $G_o$.
\end{lemma}
\begin{proof}
	Since $|T|$ is even, $T$ has a period of even length.
    Let $p_e$ be the shortest even period of $T$.
    By \Cref{lem:PreGaloisPerEven}, $U = T[1..p_e]$ is Galois if $U$ is primitive.
    Since $U$ is not primitive and $p_e$ is the smallest even period of $T$,
    we have $\OddPer(T) = p_o = p_e/2$.
    Thus, there is an odd-length prefix $G_o = T[1..p_o]$ of $U$ such that $U = G_o^2$.
\end{proof}

\begin{lemma}
\label{lemGoddsquare}
	Let $\Godd$ be an odd-length Galois root of a pre-Galois word $T = \Godd^{k}\Godd'$ with $k \ge 2$ and $\Godd'$ is a prefix of $\Godd$. 
	Then $T$ has no even-length Galois root.
\end{lemma}

\begin{proof}
	Since $T = \Godd^{k}\Godd'$, $2|\Godd|$ is a period of $T$.
    Assume that $T$ has a shorter even period $p_e < 2|\Godd|$.
    By \Cref{lem:OddBorder}, $\Godd$ does not have a proper border of even length.
		Because the two conditions (a) $\Godd \Godd$ has even length and (b) $p_e \in [|\Godd|+1..2|\Godd|-1]$
		would imply that $\Godd \Godd$ (and thus $\Godd$ due to the length of $p_e$) has a border of even length,
    $p_e$ must be less than $|\Godd|$.
		However, by the periodicity lemma (\cref{lem:periodicityLemma}),
    there exists an odd period shorter than $\Godd$,
    which contradicts that $|\Godd|$ is the shortest odd-length Galois period of $T$.
\end{proof}

\begin{example}
	Let $T = \mtt{abaa}$ be an even-length Galois word. 
	$T$ has the odd-length Galois root \mtt{aba}.
	By appending \mtt{b} to $T$, we obtain \mtt{abaab}, which is pre-Galois with no even-length Galois root.
	$T \cdot \mtt{b}$ can be written as $(\mtt{aba})^{5/3}$, a fractional repetition of \mtt{aba}.
\end{example}
By Lemmas \ref{lemGoddPrimeFull} and \ref{lemGoddsquare},
if $T$ has an even period, $T[1..p_e]$ is either $G_e$ or $G_o^2$.

\section{Determining Galois Words}\label{secDeterminingGaloisWords}
The algorithm we propose checks if a word $T$ is Galois by reading $T$ from left to right.
For that, we want to maintain the Galois roots of the prefix of $T$ read so far.
To this end, we study the Galois roots of $T' = T \cdot z$, i.e., when appending a symbol~$z$ to $T$.
Our main observation can be stated as follows:
\begin{theorem}
    Let $T$ be a pre-Galois word, $p_o = \OddPer(T)$, and $p_e = \EvenPer(T)$.
    Given a symbol $z$, the extension $T' = T \cdot z$ is a pre-Galois word if and only if
    both conditions
    $T'[1..|T| - p_o + 1] \AltPrecEq T'[p_o + 1..|T| + 1]$ and $T'[1..|T| - p_e + 1] \AltPrecEq T'[p_e + 1..|T| + 1]$ hold.
\end{theorem}

In what follows, we break down the statement of this theorem into two lemmas for each direction.
First, we consider the case where $T'$ cannot be a pre-Galois word.
\begin{lemma}
\label{lem:notGalois}
    Let $T$ be a pre-Galois word, $p_o = \OddPer(T)$, and $p_e = \EvenPer(T)$.
    Consider a symbol $z$ and the extension $T' = T \cdot z$, such that either $T'[1..|T| - p_o + 1] \AltSucc T'[p_o + 1..|T| + 1]$ or $T'[1..|T| - p_e + 1] \AltSucc T'[p_e + 1..|T| + 1]$.
    Then the extension $T'$ is not a pre-Galois word.
\end{lemma}
\begin{proof}
    We treat here only the case involving $p_o$ because the other case involving $p_e$ can be proved similarly.
    If $p_o = |T| + 1$, $T'[1..|T| - p_o + 1] = T'[p_o + 1..|T| + 1] = \varepsilon$.
    Thus, this case does not meet the requirements of the lemma statement.
    It remains to consider $p_o \le |T|$.
    For that, suppose $T'[1..|T| - p_o + 1] \AltSucc T'[p_o + 1..|T| + 1]$.
    Since $T'[1..|T| - p_o + 1] \ne T'[p_o + 1..|T| + 1]$,  $T'[p_o + 1..|T| + 1]$ not a prefix of $T'$.
    Thus, $T'$ is not pre-Galois.
\end{proof}

\begin{figure}[t]
    \centering
        \includegraphics{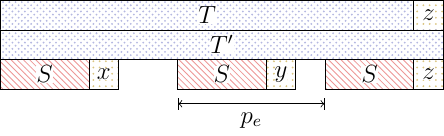}
    \caption{Sketch of the proof of \cref{lem:isPreGalois}. The caption $p_e$ can be also considered as $p_o$ for the latter case.}
    \label{fig:isPreGalois}
\end{figure}

For all other cases, we show that $T'$ is a pre-Galois word.

\begin{lemma}\label{lem:isPreGalois}
    Let $T$ be a pre-Galois word, $p_o = \OddPer(T)$, and $p_e = \EvenPer(T)$.
    Consider a symbol $z$ and the extension $T' = T \cdot z$, such that $T'[1..|T| - p_o + 1] \AltPrecEq T'[p_o + 1..|T| + 1]$ and $T'[1..|T| - p_e + 1] \AltPrecEq T'[p_e + 1..|T| + 1]$.
    Then the extension $T'$ is a pre-Galois word.
\end{lemma}
\begin{proof}
    We prove the contraposition, 
    i.e., if $T'$ is not a pre-Galois word, either $T'[1..|T| - p_o + 1] \AltSucc T'[p_o + 1..|T| + 1]$ or $T'[1..|T| - p_e + 1] \AltSucc T'[p_e + 1..|T| + 1]$ holds.

    Suppose $T'$ is not a pre-Galois word.
    Since $T$ is pre-Galois and $T'$ is not pre-Galois,
    there exists a suffix $S$ of $T$ such that $S$ is a prefix of $T$ but $S \cdot z$ is not a prefix of $T'$ and $S \cdot z \AltPrec T'$,
    i.e., $S = T'[1..|S|]$ and $S \cdot z \AltPrecM T'[1..|S|+1]$.
    In what follows we consider three cases. The first case is that $S$ is the empty word. 
    But then $z \AltPrec T'[1]$, and therefore $T'$ is not pre-Galois. The other cases concern the length-parity of $T$ and $S$.

    Consider $T$ and $S$ have the \emph{same} length-parity.
    Since $p_e$ is even, $T'[1..|T| - p_e]$ and $S$ also have the same length-parity.
    Since $S$ is a proper border of $T$, $p_e \le |T| - |S|$.
    Here we have $S = T'[1..|S|] = T'[|T| - p_e - |S|+1..|T| - p_e]$.
    Let $x = T'[|S|+1]$ and $y = T'[|T| - p_e+1]$.
    See \cref{fig:isPreGalois} for a sketch. 
    Since $|T|$ is pre-Galois, we have $S \cdot y \AltSuccEq S \cdot x$.
    Moreover, $S \cdot x \AltSucc S \cdot z$ holds by $T'[1..|S|+1] \AltSucc S \cdot z$.
    Thus we have $S \cdot y \AltSucc S \cdot z$.
    Since $T'[1..|T| - p_e]$ and $|S|$ have the same parity,
    we have $T'[1..|T| - p_e + 1] \AltSucc T'[p_e + 1..|T| + 1]$.

    Consider $T$ and $S$ have \emph{different} length-parity.
    Since $p_o$ is odd, $T'[1..|T| - p_o]$ and $|S|$ have the same parity.
    Note that $S$ is a proper border of $T$ thus $p_o \le |T| - |S|$.
    Here we have $S = T'[1..|S|] = T'[|T| - p_o - |S|+1..|T| - p_o]$.
    Let $x = T'[|S|+1]$ and $y = T'[|T| - p_o+1]$.
    Since $|T|$ is pre-Galois, we have $S \cdot y \AltSuccEq S \cdot x$.
    Moreover $S \cdot x \AltSucc S \cdot z$ holds by $T'[1..|S|+1] \AltSucc S \cdot z$.
    Thus we have $S \cdot y \AltSucc S \cdot z$.
    Since $T'[1..|T| - p_o]$ and $|S|$ have the same parity,
    we have $T'[1..|T| - p_o + 1] \AltSucc T'[p_o + 1..|T| + 1]$.
\end{proof}

Next we show how periods change when we append a symbol to a pre-Galois word $T$.
Here, we focus on $p_o$ first.
The cases for $p_e$ can be proven in a similar way.
The claim of the first lemma follows by definition.

\begin{lemma}\label{lem:oddPerSame}
    Let $T$ be a pre-Galois word and $p_o = \OddPer(T)$.
    Consider a symbol $z$ and the extension $T' = T \cdot z$, such that $z = T'[|T| - p_o + 1]$.
    Then the extension $T'$ has $\OddPer(T') = p_o$.
\end{lemma}

\begin{lemma}\label{lem:oddPerDiff}
    Let $T$ be a pre-Galois word and $p_o = \OddPer(T)$.
    Consider a symbol $z$ and the extension $T' = T \cdot z$ with $T'[1..|T| - p_o + 1] \AltPrec T'[p_o + 1..|T| + 1]$.
    Then, 
    \begin{align*}
        \OddPer(T') = 
        \begin{cases} 
            |T'|  &   \text{if } |T'| \text{~is odd,} \\
            |T'| + 1    &   \text{otherwise.}
        \end{cases}
    \end{align*}
\end{lemma}
\begin{proof}
    If $|T|$ has no odd period, i.e., $p_o = |T| + 1 = |T'|$, then $|T|$ is even.
    Thus, $|T'|$ is odd and $\OddPer(T') = |T'|$.
    Otherwise, suppose that $|T|$ has an odd period.
    Assume $T'$ has odd period $p_o' < |T'|$.
    Thus, we have $T'[1..|T| - p_o' + 1] = T'[p_o' + 1..|T| + 1]$.
    Let $S = T'[1..|T| - p_o']$ and $y = T'[|T| - p_o+1]$.
    Since $T$ is pre-Galois, we have $S \cdot z \AltPrecEq S \cdot y$.
    However, by $T'[1..|T| - p_o + 1] \AltPrec T'[p_o + 1..|T| + 1]$,
    we have $S \cdot y \AltPrec S \cdot z$, which is a contradiction.
    Therefore, $T'$ has no odd period $p_o'$ with $p_o' < |T'|$,
    which implies $\OddPer(T') = |T'|$ if $|T'|$ is odd or $\OddPer(T') = |T'| + 1$ if $|T'|$ is even.
\end{proof}

In a similar way, we can show the following lemmas.

\begin{lemma}\label{lem:evenPerSame}
    Let $T$ be a pre-Galois word and $p_e = \EvenPer(T)$.
    Consider a symbol $z$ and the extension $T' = T \cdot z$, such that $z = T'[|T| - p_e + 1]$.
    Then the extension $T'$ has $\EvenPer(T') = p_e$.
\end{lemma}

\begin{lemma}\label{lem:evenPerDiff}
    Let $T$ be a pre-Galois word and $p_e = \EvenPer(T)$. 
    Consider a symbol $z$ and the extension $T'[1..|T| - p_e + 1] \AltPrec T'[p_e + 1..|T| + 1]$.
    Then, 
    \begin{align*}
        \EvenPer(T') = 
        \begin{cases} 
            |T'|  &   \text{if } |T'| \text{ is even,} \\
            |T'| + 1    &   \text{otherwise.}
        \end{cases}
    \end{align*}
\end{lemma}

With \Cref{alg:isgalois} we give algorithmic instructions in how to verify whether an input word $T$ is Galois.
For each position in $T$, the algorithm performs a constant number of symbol comparisons on $T$.
Storing only the two periods $p_e$ and $p_o$ of the processed prefix up so far, it thus runs in linear time with a constant number of words extra to the input word $T$.
We obtain the following theorem:

\begin{algorithm2e}[t]
    \caption{Determining whether a word is Galois, see \cref{thmCheckGalois}.}
	\label{alg:isgalois}
	\SetVlineSkip{0.5mm}
	\Fn(\tcp*[f]{Assume $|T| \ge 2$, otherwise always true}){$\IsGal(T)$}{%
        $p_o = 1$;
        $p_e = 2$\;
	\For(\tcp*[f]{Loop-Invariant: $T[1..i-1]$ is pre-Galois}){$i$ \textup{\textbf{from}} $2$ \textup{\textbf{to}} $|T|$}{%
            \eIf{$i$ \textup{is odd}}{%
                \If{$p_e < i$}{
                    \lIf(\tcp*[f]{\Cref{lem:notGalois}}){$T[i] < T[i-p_e]$}{%
    		              \textbf{return} \textbf{False}%
                    }\lElseIf(\tcp*[f]{\Cref{lem:evenPerDiff}}){$T[i] > T[i-p_e]$}{%
                        $p_e = i+1$%
                    }
                }
                \If{$p_o < i$}{
                    \lIf(\tcp*[f]{\Cref{lem:oddPerDiff}}){$T[i] < T[i-p_o]$}{%
                        $p_o = i$%
                    }\lElseIf(\tcp*[f]{\Cref{lem:notGalois}}){$T[i] > T[i-p_o]$}{%
    		              \textbf{return} \textbf{False}%
                    }
                }
            }{%
                \If{$p_e < i$}{
                    \lIf(\tcp*[f]{\Cref{lem:evenPerDiff}}){$T[i] < T[i-p_e]$}{%
                        $p_e = i$%
                    }\lElseIf(\tcp*[f]{\Cref{lem:notGalois}}){$T[i] > T[i-p_e]$}{%
    		              \textbf{return} \textbf{False}%
                    }
                }
                \If{$p_o < i$}{
                    \lIf(\tcp*[f]{\Cref{lem:notGalois}}){$T[i] < T[i-p_o]$}{%
    		              \textbf{return} \textbf{False}%
                    }\lElseIf(\tcp*[f]{\Cref{lem:oddPerDiff}}){$T[i] > T[i-p_o]$}{%
                        $p_o = i+1$%
                    }
                }
            }
        }
	\uIf(\tcp*[f]{Is $T$ primitive?}){$p_o = |T|$}{%
	    \textbf{return} \textbf{True} \tcp*[f]{$T$ is Galois by \cref{lem:PreGaloisPerOdd}}
        }\ElseIf{$p_e = |T|$ \textup{and} $p_e \ne 2p_o$}{%
	    \textbf{return} \textbf{True} \tcp*[f]{$T$ is Galois by \cref{lem:PreGaloisPerEven} and \cref{lemGoddPrimeFull}.}
        }
		\textbf{return} \textbf{False} \tcp*[f]{$T$ is pre-Galois but not primitive (hence not Galois)}
	}
\end{algorithm2e}

\begin{theorem}\label{thmCheckGalois}
Given a word $T$, we can verify whether $T$ is Galois in $O(|T|)$ time with $O(1)$ working space.
\end{theorem}

\section{Computing the Galois Factorization Online}\label{secGaloisFactorization}
In this section we present an online algorithm for computing the Galois factorization of a given word.
We first start with a formal definition of the Galois factorization, introduce a key property called $\SPref(T)$, 
and then show how to compute  $\SPref(T)$.
The Galois factorization of a word $T$ is defined as follows.

\begin{definition}[Galois factorization]
    A factorization $T = G_1 \cdot G_2 \cdots G_k$ is the \emph{Galois factorization} of $T$
    if $G_i$ is Galois for $1 \le i \le k$ and $G_1 \AltSuccEq G_2 \AltSuccEq \cdots \AltSuccEq G_k$ holds.
\end{definition}

It is known that every word admits just one Galois factorization (see \cite[Th\'{e}or\`{e}me~2.1]{reutenauer05mots} or \cite{dolce19generalized}).
We denote the Galois factorization $T = G_1 \cdot G_2 \cdots G_k$ of $T$ by $\GF(T) = (G_1,G_2,\ldots,G_k)$.
The Galois factorization has the following property.

\begin{lemma}[{\cite[Theorem 33]{dolce19generalized}}]\label{lem:dolce}
    Let $\GF(T) = (G_1,G_2,\ldots,G_k)$ be the Galois factorization of a word $T$ of length~$n$.
    Let $P$ be the shortest non-empty prefix of $T$ such that
    \begin{align}\label{cond:pref}
        P \AltSuccEq T \text{ if } |P| \text{ is even } \text{  and  } P \AltPrecEq T \text{ if } |P| \text{ is odd.}
    \end{align}
    Then we have
    \begin{align*}
        P = 
        \begin{cases} 
            G_1^2  &   \text{if } |G_1| \text{ is odd, } m \text{ is even, and } m < k, \\
            G_1    &   \text{otherwise,}
        \end{cases}
    \end{align*}
    where the integer $m$ is the multiplicity of $G_1$, i.e., $G_i = G_1$ for $i \le m$, but $G_{m+1} \ne G_1$.
\end{lemma}

We denote such $P$ in \Cref{lem:dolce} by $\SPref(T)$.
If we can compute $\SPref(T)$ for any word $T$, we can compute the Galois factorization of $T$ by recursively computing $\SPref(T)$ from the suffix remaining when removing the prefix $\SPref(T)$.
To this end, we present a way to compute $\SPref(T)$ by using the periods of $T$.

\begin{figure}[t]
    \begin{minipage}{0.58\linewidth}
        \includegraphics{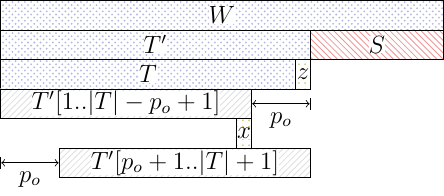}
    \end{minipage}
    \hfill
    \begin{minipage}{0.4\linewidth}
        \includegraphics{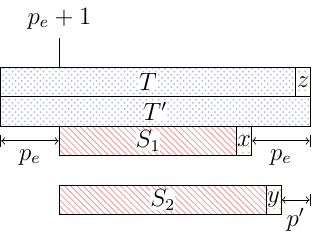}
    \end{minipage}
    \caption{Sketch of the proof of \cref{lem:factorPrefix} (left) and of \cref{lem:factorprefixeven} (right).}
    \label{fig:factorPrefix}
\end{figure}

\begin{lemma}\label{lem:factorPrefix}
    Let $T$ be a pre-Galois word, $p_o = \OddPer(T)$, and $p_e = \EvenPer(T)$.
    Consider a symbol $z$ and the extension $T' = T \cdot z$, such that
    (a) $T'[1..|T| - p_o + 1] \AltSucc T'[p_o + 1..|T| + 1]$ or 
    (b) $T'[1..|T| - p_e + 1] \AltSucc T'[p_e + 1..|T| + 1]$ hold (both (a) and (b) can hold at the same time).
    Then, for any word $S$, we have $\SPref(T'S) = T'[1..p]$ such that
    \begin{align}\label{eq:FactorPrefix}
        p = 
        \begin{cases}
        \min\{p_o,p_e\} &\text{if } T'[1..|T| - p_o + 1] \AltSucc T'[p_o + 1..|T| + 1]  \\ 
        &\text{ and } T'[1..|T| - p_e + 1] \AltSucc T'[p_e + 1..|T| + 1], \textup{~((a) and (b))}\\
        p_o &\text{if } T'[1..|T| - p_e + 1] \AltPrecEq T'[p_e + 1..|T| + 1], \textup{~((b) but not (a))}\\
    p_e &\text{otherwise. } \textup{~((a) but not (b))}
        \end{cases}
    \end{align}
\end{lemma}
\begin{proof}
Let $W=T'S$ for some arbitrary word $W \in \Sigma^*$.
Suppose $T'[1..|T| - p_o + 1] \AltSucc T'[p_o + 1..|T| + 1]$ holds.
Let $x = T'[|T| - p_o + 1]$.
Since $T$ and $T'[1..|T| - p_o]$ have different length-parities,
we have $T \cdot x \AltPrec T \cdot z$, which implies $T'[1..p_o] \AltPrec W$.
See the left of \cref{fig:factorPrefix} for a sketch.
Similarly, we can show that $T'[1..p_e] \AltPrec W$ if $T'[1..|T| - p_e + 1] \AltSucc T'[p_e + 1..|T| + 1]$.

Next, we show the minimality of $p$.
Consider a prefix $X$ such that $|X| < p$.
Let $T[1..q]$ be the longest prefix of $T$ such that $|X|$ is its period.
Then, we have $T[1..q - |X|] = T'[|X| + 1..q]$ and $T[1..q - |X| + 1] \ne T'[|X| + 1..q + 1]$.
Since $T$ is pre-Galois, we have $T[1..q - |X| + 1] \AltPrec T'[|X| + 1..q + 1]$.
If $|X|$ is odd, $X \AltSucc T[1..q  + 1]$, 
otherwise if $|X|$ is even, $X \AltPrec T[1..q  + 1]$, which does not satisfy the condition of \cref{cond:pref} in \cref{lem:dolce}.
\end{proof}

By using the property shown in \Cref{lem:factorPrefix},
we can compute $\GF(W)$ by computing $\SPref(W)$ recursively.
For example, given $W = UV$ with $U = \SPref(W)$,
after computing $\SPref(W)$ to get $U$, we recurse on the remaining suffix $V$ and compute $\SPref(V)$ to get the next factor.
We can modify \Cref{alg:isgalois} to output $\SPref(W)$ in $O(\ell)$ time, where $\ell$ is the length of the longest pre-Galois prefix of $W$.
However, it takes time if we compute $\GF(W)$ by the recursive procedure,
especially when $\ell$ is much larger than $|\SPref(W)|$.
To tackle this problem, we use the following property.

\begin{lemma}
\label{lem:factorprefixeven}
    Let $T$ be a pre-Galois word and $p_e = \EvenPer(T)$.
    Consider a symbol $z$ and the extension $T' = T \cdot z$, such that $T'[1..|T| - p_e + 1] \AltSucc T'[p_e + 1..|T| + 1]$.
    Let $\SPref(T') =  T'[1..p]$.
    If $p = p_e$ and $|T| \ge 2p$, we have $\SPref(T'[p+1..]) = T'[p+1..2p] = T'[1..p]$.
\end{lemma}
\begin{proof}
   Since $T'[1..|T| - p_e]$ and $T'[p_e+1..|T| - p_e]$ have the same length-parity,
   we have $T'[p_e+1..|T| - p_e + 1] \AltSucc T'[2p_e + 1..|T| + 1]$.
   Assume that $T[p_e+1..]$ has a period $p' < p_e$ and $T'[p_e+1..|T| - p' + 1] \AltSucc T'[p_e + p' + 1..|T| + 1]$.
   Since $T$ is pre-Galois, $T'[1..p_e] = T'[p_e + 1..2p_e]$ has no even border.
   Thus $p'$ is odd.
   Let $S_1$ = $T'[p_e+1..|T| - p_e]$, $S_2$ = $T'[p_e+1..|T| - p']$, $x = T'[|T| - p_e + 1]$, and $y = T'[|T| - p' + 1]$.
   By $T'[p_e+1..|T| - p' + 1] \AltSucc T'[p_e + p' + 1..|T| + 1]$, 
   we have $S_2\cdot y \AltSucc S_2\cdot z$.
   See the right of \cref{fig:factorPrefix} for a sketch.
   Since $S_1$ and $S_2$ have different parities, 
   we have $S_1\cdot y \AltPrec S_1\cdot z$.
   Moreover, by $T'[p_e+1..|T| - p_e + 1] \AltSucc T'[2p_e + 1..|T| + 1]$, we have $S_1\cdot x \AltSucc S_1\cdot z$,
   which implies $S_1\cdot x \AltSucc S_1\cdot y$.
   However, since $T[p_e + 1..]$ is pre-Galois,
   $S_1\cdot x \AltPrecEq S_1\cdot y$, which contradicts $S_1\cdot x \AltSucc S_1\cdot y$.
   Therefore, $T[p_e+1..]$ has no period $p'$ with $p' < p_e$.
\end{proof}

Let $U = \SPref(W)$, $|U|$ is even, and $W = U^kV $ for some $k \ge 2$.
By \Cref{lem:factorprefixeven}, we know that $\SPref(U^jV) = U$ for $1 \le j < k$ without computing it explicitly.
Next, we consider the case when $|\SPref(W)|$ is odd.

\begin{lemma}
\label{lem:factorprefixodd}
    Let $T$ be a pre-Galois word, $p_o = \OddPer(T)$, and $p_e = \EvenPer(T)$.
    Consider a symbol $z$ and the extension $T' = T \cdot z$, such that $T'[1..|T| - p_o + 1] \AltSucc T'[p_o + 1..|T| + 1]$.
    Let $P = T'[1..p] = \SPref(T')$.
    If $p = p_o$ and $|T| \ge 3p$, we have $\SPref(T'[p+1..]) = T'[p+1..3p] = T'[1..2p]$.
\end{lemma}

\begin{proof}
    Since $T$ is pre-Galois, $T'[1..p_o] = T'[p_o + 1..2p_o]$ does not have an even border.
    Thus $\Per_o(T[p_o+1..]) = p_o$ and $\Per_e(T[p_o+1..]) = 2p_o$.
    Moreover, since $T'[1..|T| - p_o]$ and $T'[p_o+1..|T| - p_o]$ have different parities,
    we have $T'[p_o+1..|T| - p_o + 1] \AltPrec T'[2p_o + 1..|T| + 1]$.
    Next, $T'[p_o+1..|T| - 2p_o + 1] \AltSucc T'[3p_o + 1..|T| + 1]$ holds, since $T'[1..|T| - p_o]$ and $T'[p_o+1..|T| - 2p_o]$ have the same parity.
    Therefore, $\SPref(T'[p_o+1..]) = T'[p_o+1..3p] = T'[1..2p_o]$.
\end{proof}

\begin{algorithm2e}[t!]
    \caption{Computing the Galois factorization, as claimed in \cref{thmGaloisFactorization}.}
	\label{alg:galoisfactorization}
	\SetVlineSkip{0.5mm}
	\Fn{$\GFact(T)$}{%
        Append $\$$ to $T$\;
        $\Fact = (\ )$ empty list;
        $i = 0$\;
        \While{$i \le |T|$}{\label{line:outerLoop}
            $p_o = 1$;
            $p_e = 2$\;
            \For{$j$ \textup{from} $2$ \textup{to} $|T| - i$}{\label{line:innerLoop}
                $p = |T|+2$;
                $p_e' = p_e$;
                $p_o' = p_o$\;
                \eIf{$j$ \textup{is odd}}{%
                    \If{$p_e < j$}{
                        \lIf(\tcp*[f]{\Cref{lem:factorPrefix}}){$T[i+j] < T[i+j-p_e]$}{%
        		              $p = \min\{p,p_e\}$%
                        }\lElseIf(\tcp*[f]{\Cref{lem:evenPerDiff}}){$T[i+j] > T[i+j-p_e]$}{%
                            $p_e' = j+1$%
                        }
                    }
                    \If{$p_o < j$}{
                        \lIf(\tcp*[f]{\Cref{lem:oddPerDiff}}){$T[i+j] < T[i+j-p_o]$}{%
                            $p_o' = j$%
                        }\lElseIf(\tcp*[f]{\Cref{lem:factorPrefix}}){$T[i+j] > T[i+j-p_o]$}{%
        		              $p = \min\{p,p_o\}$%
                        }
                    }
                }{%
                    \If{$p_e < j$}{
                        \lIf(\tcp*[f]{\Cref{lem:evenPerDiff}}){$T[i+j] < T[i+j-p_e]$}{%
                            $p_e' = j$%
                        }\lElseIf(\tcp*[f]{\Cref{lem:factorPrefix}}){$T[i+j] > T[i+j-p_e]$}{%
        		              $p = \min\{p,p_e\}$%
                        }
                    }
                    \If{$p_o < j$}{
                        \lIf(\tcp*[f]{\Cref{lem:factorPrefix}}){$T[i+j] < T[i+j-p_o]$}{%
        		              $p = \min\{p,p_o\}$%
                        }\lElseIf(\tcp*[f]{\Cref{lem:oddPerDiff}}){$T[i+j] > T[i+j-p_o]$}{%
                            $p_o' = j+1$%
                        }
                    }
                }
                \If{$p \ne |T|+2$}{\label{line:factorPrefix2}
                    \While{$j > p$}{%
                        \uIf(\tcp*[f]{\Cref{lem:factorprefixeven}}){$p = p_e$ \textup{and} $p_e = 2p_o$}{%
                            Append $T[i..i+p_o-1]$ and $T[i+p_o..i+2p_o-1]$ to $\Fact$\;
                        }\Else{
                            Append $T[i..i+p-1]$ to $\Fact$\;
                        }
                        $i = i + p$;
                        $j = j - p$\;
                        $p = p_e$\;%
                    }
                    \textbf{break}\;
                }
                $p_e = p_e'$;
                $p_o = p_o'$\;
            }
        }
		\textbf{return} $\Fact$\;
	}
\end{algorithm2e}

Let $U = \SPref(W)$, $|U|$ is odd, and $W = U^kV $ for some $k \ge 3$.
By Lemmas \ref{lem:factorprefixeven} and \ref{lem:factorprefixodd},
we know that $\SPref(U^{k-2j-1}V) = U^2$ for $0 \le j < \lceil k/2 \rceil$ without computing it explicitly.

Lemmas~\ref{lem:factorPrefix}, \ref{lem:factorprefixeven}, and \ref{lem:factorprefixodd} are used to factorize a pre-Galois word when we extended it.
However, we can not use the Lemmas to factorize $T$ when $T$ is pre-Galois but not Galois and the input is terminated.
Although we can factorize $T$ by finding $P = \SPref(T)$ in \Cref{lem:dolce},
we need an additional procedure to find such $P$.
To simplify our algorithm, we append a terminal symbol $\$$ that is smaller than all symbols of $\Sigma$. In particular, all other symbols in $W$ are different from $\$$.\footnote{%
Without $\$$, the last factor we report might be just pre-Galois, not Galois. 
So we have to break the last factor into Galois factors. 
If $W$ ends with the unique symbol $\$$, then $\$$ cannot be included in another Galois factor of $W$;
it has to stay alone as a Galois factor of length one, and thus we cannot end with the last factor being just pre-Galois.}
Here we show that the appended $\$$ determines a Galois factor of length one, thus it does not affect the factorization result.

\begin{lemma}
\label{lem:withterminal}
    Consider a symbol $\$$ that does not appear in a word $T$ and $\$ \prec c$ for any $c \in \Sigma$.
    Then, $\GF(T) = (G_1, G_2, \dots, G_k)$ iff $\GF(T \cdot \$) = (G_1, G_2, \dots, G_k, \$)$.
\end{lemma}
\begin{proof}
    Let $\GF(T) = (G_1, G_2, \dots, G_k)$.
    Here, $G_1 \AltSuccEq  G_2 \AltSuccEq \dots \AltSuccEq G_k$.
    Since $\$ \prec c$ for any $c \in \Sigma$,
    we have $G_1 \AltSuccEq  G_2 \AltSuccEq \dots \AltSuccEq G_k \AltSuccEq \$$.
    Therefore, $\GF(T \cdot \$) = (G_1, G_2, \dots, G_k, \$)$.
    Similarly, let $\GF(T \cdot \$) = (G_1, G_2, \dots, G_k, \$)$.
    Here, $G_1 \AltSuccEq  G_2 \AltSuccEq \dots \AltSuccEq G_k \AltSuccEq \$$.
    Therefore, $\GF(T) = (G_1, G_2, \dots, G_k)$.
\end{proof}

This gives us the final ingredient for introducing the algorithmic steps for computing the Galois factorization, which we present as pseudo code in \cref{alg:galoisfactorization}.

\begin{theorem}\label{thmGaloisFactorization}
    The Galois factorization of a word $T$ can be computed in $O(|T|)$ time and $O(1)$ additional working space, excluding output space.
\end{theorem}
\begin{proof}
    The correctness of \Cref{alg:galoisfactorization} is proven by Lemmas~\ref{lem:factorPrefix}, \ref{lem:factorprefixeven}, \ref{lem:factorprefixodd}, and \ref{lem:withterminal}.
    Next, we prove the time complexity of \Cref{alg:galoisfactorization}.
    The time complexity of the algorithm is bounded by the number of iterations of the inner loop (Line~\ref{line:innerLoop}).
    The algorithm increments $j$ until it finds a prefix to be factorized (Line~\ref{line:factorPrefix2}).
    Here we show that $j \le 3\ell + 1$, where $\ell$ is the length of the factorized prefix.
    Let $p = p_e$.
    If $j < 2p+1$, it is clear that $j \le 3\ell + 1$, where $\ell = p$.
    Otherwise, if $j \ge 2p+1$, the algorithm factorizes the prefix recursively $k$ times,
    such that $kp \ge j$ and $(k+1)p > j$.
    Thus, we have $\ell = kp$ which implies $j \le 3\ell + 1$.
    On the other hand, let $p = p_o$.
    If $j < 3p+1$, its clear that $j \le 3\ell + 1$, where $\ell = p$.
    Otherwise, if $j \ge 3p+1$, the algorithm factorizes the prefix recursively $k$ times,
    such that $kp \ge j$ and $(k+2)p > j$.
    Thus, we have $\ell = kp$ which implies $j \le 3\ell + 1$.
    Therefore, the number of iterations of the inner loop is $O(|T|)$, since the total length of the factors is $|T|$.
\end{proof}

\section{Computing Galois rotation}\label{secGaloisRotation}
While we can infer the Lyndon rotation of a word $T$ from the Lyndon factorization of $T\cdot T$, the same kind of inference surprisingly does not work for Galois words~\cite[Example 41]{dolce19generalized}.
Here, we present an algorithm computing the Galois rotation, using constant additional working space.
The algorithm is a modification of  \Cref{alg:galoisfactorization}.
We start with formally defining the Galois rotation of a word.

\begin{definition}
    Let $W$ be a primitive word.
    A rotation $T = VU$ is a \emph{Galois rotation} of $W = UV$ if $T$ is Galois.
\end{definition}

To describe our algorithm computing Galois rotations,
we study a property of the Galois factorization for repetitions of a Galois word.

\begin{algorithm2e}[!t]
    \caption{Computing the Galois rotation of $T$, as claimed in \cref{thm:galoisRotation}.}
	\label{alg:galoisRotation}
	\SetVlineSkip{0.5mm}
	\Fn{$\GRot(T)$}{%
        $i = 0$\;
        \While{$i \le 3|T|$}{
            $p_o = 1$;
            $p_e = 2$\;
            \For{$j$ \textup{from} $2$ \textup{to} $3|T| - i$}{\label{line:RotationInnerLoop}
		$p = 3|T|+2$;
                $p_e' = p_e$;
                $p_o' = p_o$\;
                \eIf{$j$ \textup{is odd}}{%
                    \If{$p_e < j$}{
                        \lIf(\tcp*[f]{\Cref{lem:factorPrefix}}){$T[i+j] < T[i+j-p_e]$}{%
        		              $p = \min\{p,p_e\}$%
                        }\lElseIf(\tcp*[f]{\Cref{lem:evenPerDiff}}){$T[i+j] > T[i+j-p_e]$}{%
                            $p_e' = j+1$%
                        }
                    }
                    \If{$p_o < j$}{
                        \lIf(\tcp*[f]{\Cref{lem:oddPerDiff}}){$T[i+j] < T[i+j-p_o]$}{%
                            $p_o' = j$%
                        }\lElseIf(\tcp*[f]{\Cref{lem:factorPrefix}}){$T[i+j] > T[i+j-p_o]$}{%
        		              $p = \min\{p,p_o\}$%
                        }
                    }
                }{%
                    \If{$p_e < j$}{
                        \lIf(\tcp*[f]{\Cref{lem:evenPerDiff}}){$T[i+j] < T[i+j-p_e]$}{%
                            $p_e' = j$%
                        }\lElseIf(\tcp*[f]{\Cref{lem:factorPrefix}}){$T[i+j] > T[i+j-p_e]$}{%
        		              $p = \min\{p,p_e\}$%
                        }
                    }
                    \If{$p_o < j$}{
                        \lIf(\tcp*[f]{\Cref{lem:factorPrefix}}){$T[i+j] < T[i+j-p_o]$}{%
        		              $p = \min\{p,p_o\}$%
                        }\lElseIf(\tcp*[f]{\Cref{lem:oddPerDiff}}){$T[i+j] > T[i+j-p_o]$}{%
                            $p_o' = j+1$%
                        }
                    }
                    
                }
                \If{$p \ne 3|T|+2$}{

                    \While{$j > p$}{%
                        $i = i + p$;
                        $j = j - p$\;
                        $p = p_e$\;
                    }
                    \textbf{break}\;
                }
                $p_e = p_e'$;
                $p_o = p_o'$\;
                \If{$p_o \ge |T|$ \textup{and} $p_e \ge |T|$}{
                    \textbf{return} $(i\mod |T|)+1$\;
                }
            }
        }
	}
\end{algorithm2e}

\begin{lemma}
\label{lem:GaloisSquare}
    Let $T$ be a Galois word and $P = \SPref(T^k)$ for some rational number $k \ge 2$.
    Then $|P| \ge |T|$.
\end{lemma}
\begin{proof}
    Suppose that $|P| < |T|$ and $|P|$ is even.
    Since $T$ is primitive, 
    $|P|$ is not a period of $T^2$ by \Cref{lem:periodicityLemma}.
    Thus, there exist a position $i < 2|T|$ such that $P^\omega[i] \ne T^2[i]$.
    Moreover, we have $P \AltSuccEq T^k$ by \Cref{lem:dolce}, which implies $P \AltSucc T^2$. 
    However, we have $P \AltPrec T \AltEq T^2$ by \Cref{lem:dolceGalois}, which is a contradiction.
    The case that $|P|$ is odd leads to a similar contradiction, and thus $|P| \ge |T|$ must hold.
\end{proof}

We then use the above property to show the following lemma, which is the core of our algorithm.
\begin{lemma}\label{lem:GaloisRot}
    Let $W$ be the Galois rotation of a primitive word $T$.
    Given $TTT = UWWV$ with $|U| < |T|$,
    let $\GF(U) = (G_1,G_2,\dots,G_k)$ and $\GF(WWV) = (H_1,H_2,\dots,H_l)$.
    Then we have $\GF(UWWV) = (G_1,G_2,\dots, G_k,H_1,H_2,\dots,H_l)$.
\end{lemma}
\begin{proof}
    For $U = \varepsilon$, the claim trivially holds with $V = W$ and
    $\GF(UWWV) = \GF(WWV) = (H_1,H_2,\dots,H_l)$.
    In the rest of the proof we assume $U \ne \varepsilon$.
    Let $\GF(U) = (G_1,G_2,\dots,G_k)$ and $\GF(WWV) = (H_1,H_2,\dots,H_l)$.
    Because $W = V \cdot U$, $U$ (and in particular $G_k$) is a suffix of $W$.
    By the definition of the Galois factorization, we have $G_1 \AltSuccEq G_2 \AltSuccEq \dots \AltSuccEq G_k$ and $H_1 \AltSuccEq H_2 \AltSuccEq \dots \AltSuccEq H_l$.
    By showing $G_k \AltSuccEq H_1$, we obtain that the factorization $(G_1,G_2,\dots, G_k,H_1,H_2,\dots,H_l)$ of $UWWV$ admits the properties of the Galois factorization, which proves the claim.
    
    To that end, we first observe that $G_k$ is a proper suffix of $W$ and $W$ is Galois,
    thus $G_k \AltSucc W$.
    Next, if $G_k$ is not a prefix of $W$,
    there exists a position $i \le |G_k| < |W|$ such that $G_k^\omega[i] \ne W[i]$.
    Otherwise if $G_k$ is a prefix of $W$,
	$G_k$ is a border of $W$ and $|W|-|G_k|$ is a period of $W$.
    Assuming that $|G_k|$ is a period of $W$, 
    the greatest common divisor $\mathsf{gcd}$ of $|G_k|$ and $|W|-|G_k|$ is a period of $G_k$ by the periodicity lemma (\cref{lem:periodicityLemma}), and $\mathsf{gcd}$ is a factor of $|W|$.
    However this is impossible since $W$ is primitive; thus $|G_k|$ cannot be a period of $W$.
    Hence, there exists a position $i \le |W|$ such that $G_k^\omega[i] \ne W[i]$.
    We therefore know that the first $k$ Galois factors in $\GF(U)$ and $\GF(UWWV)$ are the same since we cannot extend $G_k$ further without losing the property to be Galois.
    Moreover, $W$ is a prefix of $H_1$ by \Cref{lem:GaloisSquare}.
    Thus, there exists a position $j \le |W| \le |H_1|$ such that $G_k^\omega[j] \ne H_1[j]$,    
    which implies $G_k \AltSucc H_1$.
    Therefore, we have $G_1 \AltSuccEq G_2 \AltSuccEq \dots \AltSuccEq G_k \AltSuccEq H_1 \AltSuccEq H_2 \AltSuccEq \dots \AltSuccEq H_l$, which implies $\GF(UWWV) = (G_1,G_2,\dots, G_k,H_1,H_2,\dots,H_l)$.
\end{proof}

\begin{table}[!t]
	\centering
	\caption{Counting the number of Galois factors for various datasets.
    The alphabet size is denoted by $\sigma$.
     Counts are listed in the \#~columns, together with a time evaluation with Duval's algorithm computing the Lyndon factorization.
 \emph{Upper part:} The Canterbury and Calgary corpus datasets.
 \emph{Lower part:} The Pizza\&Chili corpus datasets.
 Note that we used different time units for the upper table (microseconds) and lower table (seconds).
 }
	\label{tab:galoisFactors}
\begin{tabular}{l*{6}{r}}
    \toprule
		     & & & \multicolumn{2}{c}{Galois} & \multicolumn{2}{c}{Lyndon} \\
	\cmidrule(lr){4-5}
	\cmidrule(lr){6-7}
	file & $\sigma$ & size [KB] & \# & time [\textmu s] & \#  & time [\textmu s]\\
    \midrule
   \textsc{alice29.txt}          &       \num{74} & \num{152} & \num{14} & \num{3070} & \num{3} & \num{192}  \\
\textsc{asyoulik.txt}         &          \num{68} & \num{125} & \num{7} & \num{2435} & \num{2} & \num{134}  \\
\textsc{bib}          &                  \num{81} & \num{111} & \num{25} & \num{2372} & \num{6} & \num{110}   \\
\textsc{book2}          &                \num{96} & \num{610} & \num{20} & \num{12182} & \num{27} & \num{555}  \\
\textsc{cp.html}          &              \num{86} & \num{24} & \num{7} & \num{544} & \num{8} & \num{21}   \\
\textsc{fields.c}          &             \num{90} & \num{11} & \num{18} & \num{237} & \num{13} & \num{12}   \\
\textsc{grammar.lsp}          &          \num{76} & \num{3} & \num{10} & \num{78} & \num{8} & \num{6}   \\
\textsc{lcet10.txt}          &           \num{84} & \num{426} & \num{12} & \num{8599} & \num{6} & \num{438}   \\
\textsc{news}        &                   \num{98} & \num{377} & \num{24} & \num{7440} & \num{24} & \num{375}   \\
\textsc{paper1}          &               \num{95} & \num{53} & \num{19} & \num{1016} & \num{9} & \num{40}   \\
\textsc{paper2}          &               \num{91} & \num{82} & \num{14} & \num{1593} & \num{16} & \num{66}   \\
\textsc{paper3}          &               \num{84} & \num{46} & \num{11} & \num{908} & \num{14} & \num{39}   \\
\textsc{paper4}          &               \num{80} & \num{13} & \num{8} & \num{267} & \num{6} & \num{12}   \\
\textsc{paper5}          &               \num{91} & \num{11} & \num{9} & \num{237} & \num{6} & \num{10}   \\
\textsc{paper6}          &               \num{93} & \num{38} & \num{12} & \num{740} & \num{15} & \num{32}   \\
\textsc{plrabn12.txt}    &               \num{81} & \num{481} & \num{4} & \num{9801} & \num{6} & \num{559}   \\
\textsc{progc}          &                \num{92} & \num{39} & \num{15} & \num{788} & \num{12} & \num{36}   \\
\textsc{progl}          &                \num{87} & \num{71} & \num{84} & \num{1423} & \num{77} & \num{63}   \\
\textsc{progp}          &                \num{89} & \num{49} & \num{14} & \num{944} & \num{12} & \num{38}   \\
\textsc{xargs.1}         &               \num{74} & \num{4} & \num{6} & \num{88} & \num{9} & \num{5}   \\
     \midrule
     \midrule


		     & & & \multicolumn{2}{c}{Galois} & \multicolumn{2}{c}{Lyndon} \\
 	\cmidrule(lr){4-5}
 	\cmidrule(lr){6-7}
 	file & $\sigma$ & size [KB] & \# & time [s] & \#  & time [s]\\
     \midrule
	\textsc{dblp.xml} & \num{97} & \num{296135} & \num{3} & \num{5.969}  & \num{15} & \num{0.294}\\
\textsc{dna     } & \num{97} & \num{403927} & \num{26} & \num{7.799}  & \num{18} & \num{0.360}\\
\textsc{proteins} & \num{27} & \num{1184051} & \num{29} & \num{24.384}  & \num{30} & \num{1.091}\\
\textsc{sources } & \num{230} & \num{210866} & \num{23} & \num{4.307}  & \num{35} & \num{0.179}\\
    \bottomrule
\end{tabular}
\end{table}

With \cref{lem:GaloisRot}, we now have a tool to find the Galois rotation $W$ of $T$ in $TTT$ by determining $H_1$ and knowing that $W$ is a prefix of $H_1$. Since we can write $TTT = UWWV$ with $W = VU$ and $|U| < |T|$, the goal is to determine $U$. For $U$, we know that all its Galois factors have even or odd periods shorter than $|T|$, so it suffices to find the first Galois factor in $TTT$ for which both periods are at least $|T|$ long (cf.~\cref{lem:GaloisSquare}).

In detail, let $G$ be the first factor of $\GF(TTT)$ with $|G| \ge T$.
From \Cref{lem:GaloisRot} we know that the Galois rotation $W$ of a word $T$ is the prefix of $G$ whose length is $|T|$, i.e. $W = G[1..|T|]$.
\Cref{alg:galoisRotation} describes an algorithm to compute the Galois rotation of an input word $T$.
The algorithm scans $TTT$ sequentially from the beginning, mimicking our Galois factorization algorithm, except that it does not output the factors.
At the time where we set $p_o \ge |T|$ and $p_e \ge |T|$,
we know that the length of next factor we compute is $|T|$ or longer.
At that time, we can determine $G$.
To this end, we output the starting position $i$ of $G$ when reaching the condition that $p_o \ge |T|$ and $p_e \ge |T|$. 
In a post-processing, we determine $W = (TTT)[i..i+|T|-1]$, where $W$ is the Galois rotation of $T$.
To keep the additional working space constant, we do not load three copies of $T$ into memory, but use that fact that $(TTT)[k] = T[((k-1) \mod |T|) + 1]$ for $k > |T|$ when processing the input $TTT$.

\begin{theorem}\label{thm:galoisRotation}
    The Galois rotation of a word $T$ can be computed in $O(|T|)$ time and $O(1)$ additional working space.
\end{theorem}

\section{Experiments}

We have implemented our algorithms in C++.
The software is freely available by the link on the title page.
For a short demonstration, we computed the Galois factors of files from the Canterbury, the Calgary~\cite{bell89modeling} and the Pizza\&Chili corpus~\cite{ferragina08compressed},
and depict the results in \cref{tab:galoisFactors}.
We have omitted those files that contain a zero byte, which is prohibited in our implementation.
The experiments run on WSL with Intel Core i7-10700K CPU.
To compare the time with a standard Lyndon factorization algorithm, we used the implementation of Duval's algorithm from
\url{https://github.com/cp-algorithms/cp-algorithms}.

\bibliographystyle{plainnat}

\bibliography{references}

\end{document}